\documentclass[12pt]{iopart}
\usepackage[dvips]{graphicx}
\newcommand{\De}{\Delta}
\newcommand{\de}{\delta}
\newcommand{\eps}{\epsilon}
\newcommand{\la}{\lambda}
\newcommand{\si}{\sigma}
\newcommand{\Si}{\Sigma}
\newcommand{\pa}{\partial} 
\newcommand{\pla}{p_{\lambda}}
\newcommand{\minuit}{{\sc minuit}}
\newcommand{\epdflib}{{\sc epdflib}}
\newcommand{\avg}[1]{\ensuremath{\langle #1 \rangle}}
\begin{document}

\begin{flushright}
\tt\normalsize{NIKHEF-01-014}\\
\end{flushright}
\vspace{1cm}
\title{Error Estimates on Parton Density Distributions}
\author{M Botje}
\address{NIKHEF, PO Box 41882, 1009DB Amsterdam, The Netherlands}
\ead{m.botje@nikhef.nl}

\begin{abstract}
Error estimates on parton density distributions are presently based on
the traditional method of least squares minimisation and linear error
propagation in global QCD fits. We review the underlying assumptions
and the various mathematical representations of the method and address
some technical issues encountered in such a global analysis. Parton
distribution sets which contain error information are described.
\end{abstract}

\maketitle


\section{Introduction}\label{se:intro}

Parton density distributions are important ingredients in the calculation
of hard scattering lepton-hadron and hadron-hadron cross sections.  To
judge the comparison of theory and experiment it is important to
evaluate the experimental and theoretical uncertainties on the
predicted cross sections which are often dominated by those on the
input parton densities.  These densities are non-perturbative
and are obtained from QCD fits to a large body of scattering
data.  The errors which have to be propagated in such a fit are the
statistical and correlated systematic errors on the measurements,
errors on the input parameters (flavour thresholds, $\alpha_s$ etc.)
and uncertainties in the theoretical modelling (scale errors,
non-perturbative contributions like higher twists and nuclear effects,
functional form of the parton densities and so on).

In the past the errors associated with the parton densities were often
determined from the spread between different parton distribution
sets. Such a spread is of course by no means a representation of the
experimental and theoretical uncertainty. In recent years considerable
progress is made on the proper error propagation in global QCD fits
because the increasing accuracy of the HERA and Tevatron data demands
reliable estimates of the uncertainties in the theory predictions.
Parton density uncertainties are of course also important for the
ongoing LHC simulation studies.

Two methods of error propagation are presently in use.  One is based
on the method of statistical inference using  Monte Carlo integration
techniques~\cite{ref:giele}.   The more conventional approach makes
use of the standard methods of least squares minimisation and
linear error propagation. In this report we will restrict ourselves to
a description of the latter method because it is, at present, the most
developed and widely used in QCD fits by the various experimental
collaborations and theory groups.

\section{Least Squares Minimisation}\label{se:minimize}

The justification for using least squares lies in the assumption that
the measurement errors are Gaussian distributed.  To introduce the
effect of point to point correlated systematic errors,  the
measurements ($\bi{m}$) are related to the theory prediction
($\bi{t}$) by
\begin{equation}\label{eq:midef}
   m_i = t_i(\bi{p}) + r_i \si_i + \sum_k s_k \De_{ik}
\end{equation}
where $m_i$ is the measurement of data point $i$, $t_i(\bi{p})$ is the
model prediction depending on a set of parameters $\bi{p}$,  $\si_i$
is the uncorrelated (statistical) error on data point $i$ and
$\De_{ik}$ is the correlated (systematic) error from  source
$k$.

In (\ref{eq:midef}), $r_i$ and $s_k$ denote Gaussian random variables
with zero mean and unit variance. These random variables are assumed
to be independent of each other:
\begin{equation}\label{eq:delridelrj}
  \avg{\De r_i \De r_j} = \avg{\De s_i \De s_j} = \de_{ij} \qquad 
  \avg{\De r_i \De s_j} = 0
\end{equation}
where we have introduced the notation $\De x = x - \avg{x}$ and
where  the symbol $\avg{\ }$ denotes an average. The probability
density function of the measurements can then be written as a
multivariate Gaussian distribution
\begin{equation}\label{eq:dpdef}
   P = C \exp(-\case12 \chi^2)
\end{equation}
with $C$ a normalisation constant and $\chi^2$  defined by
\begin{equation}\label{eq:chi2def}
   \chi^2 = \sum_{ij} (m_i-t_i) V_{ij}^{-1} (m_j-t_j).
\end{equation}
From (\ref{eq:midef}) and (\ref{eq:delridelrj}) the
covariance matrix $\bi{V}$ of the measurements is given by
\begin{equation}\label{eq:vmeasdef}
  V_{ij} = \avg{\De m_i \De m_j} = 
  \delta_{ij}\si_i^2 + \sum_k \De_{ik} \De_{jk}.
\end{equation}
Optimal values for the parameters, $\bi{p}_0$, are found by maximising
the probability density $P$ or, equivalently, by minimising $\chi^2$
defined by (\ref{eq:chi2def}).

The standard method to calculate the covariance matrix of the fitted
parameters is based on the assumption that the theory prediction
varies approximately linearly with $\bi{p}$ near the minimum of
$\chi^2$ at $\bi{p}_0$. Expanding $\chi^2$ up to second order in the
variation $\De \bi{p} = \bi{p}-\bi{p}_0$ gives
\begin{equation}\label{eq:expandchi}
  \De \chi^2 = \chi^2(\bi{p}) - \chi^2(\bi{p}_0) =
  \sum_i \frac{\pa \chi^2}{\pa p_i}\; \De p_i + \sum_{ij} \frac{1}{2}
  \frac{\pa^2 \chi^2}{\pa p_i \pa p_j}\; \De p_i \De p_j.
\end{equation}
Because the linear term in (\ref{eq:expandchi}) vanishes at the
minimum we find for the covariance matrix $\bi{V}_p$ of the parameters
\begin{equation}\label{eq:hessian}
  V_{ij}^p = \avg{\De p_i \De p_j} = 
  \De \chi^2 \left( \frac{1}{2} \frac{\pa^2 \chi^2}{\pa p_i \pa p_j}
  \right)^{-1} =
  \De \chi^2 \; H^{-1}_{ij}
\end{equation}
where we have introduced the Hessian matrix $\bi{H}$ of second order
derivatives.  Notice that the covariance matrix $\bi{V}_p$ depends on
the choice of $\De \chi^2$ which usually, but not always, is taken to
be $\De \chi^2 = 1$. With this choice the probability density given in
(\ref{eq:dpdef}) drops by a factor $\sqrt e$ when the parameters
$\bi{p}$ are a distance $\De \bi{p}$ away from the optimum. This
corresponds to the definition of the width of a Gaussian distribution.

Having obtained the best values and the covariance matrix of the
parameters, the covariance of any two functions $F(\bi{p})$ and
$G(\bi{p})$ can be calculated with the standard formula for linear
error propagation
\begin{equation}\label{eq:fgmat}  
  \avg{\De F \De G} = \sum_{ij} \frac{\pa F}{\pa p_i}\; V_{ij}^p\;
  \frac{\pa G}{\pa p_j}
  = \De \chi^2 \sum_{ij} \frac{\pa F}{\pa p_i}\; H_{ij}^{-1}\;
  \frac{\pa G}{\pa p_j}.
\end{equation}
Here it is assumed that both $F$ and $G$ vary approximately linearly
with $\bi{p}$ in the neighbourhood of $\bi{p}_0$. Because $\avg{\De F^2} >
0$ it follows from (\ref{eq:fgmat}) that the Hessian (and its inverse)
must be positive definite, that is, for arbitrary vectors $\bi{X}$,
\begin{equation}\label{eq:posdef}
  \bi{X H X} > 0.
\end{equation}

Finally we remark that, for simplicity, we have tacitly assumed that
all errors can be treated as offsets. Notice that multiplicative
(normalisation) errors must be treated somewhat
differently~\cite{ref:agostini} to avoid possible large biases in the
fit results~\cite{ref:takeuchi}.


\section{The Hessian Matrix}\label{se:hessianmatrix}

\begin{figure}[tbh]
\begin{center}
\includegraphics[width=0.8\linewidth]{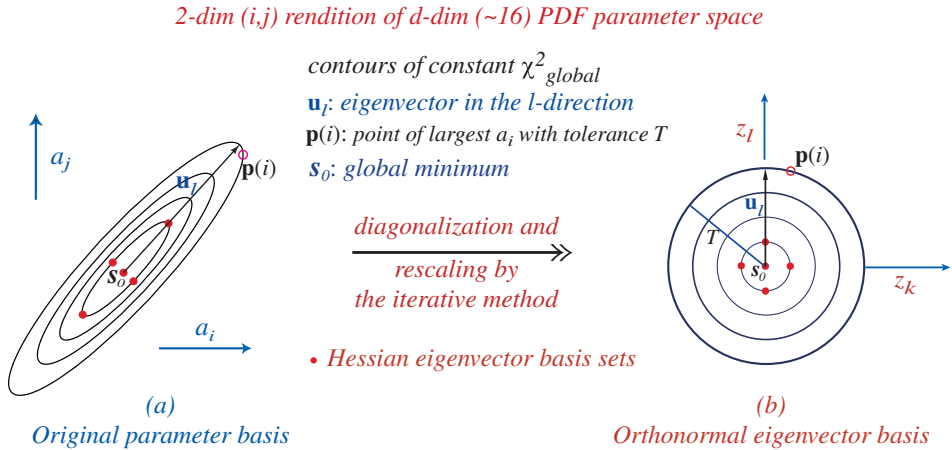}
\end{center}
\caption{\label{fig:hessecontour}
(a) Ellipsoid contours defined by constant values of $\chi^2$ in
parameter space. (b) Hyper-sphere contours of constant $\chi^2$
after an orthogonal transformation of the parameters defined by the 
eigenvectors and eigenvalues of the Hessian matrix. 
Figure taken from~\cite{ref:hmethod}.
}
\end{figure}

In the quadratic approximation (\ref{eq:expandchi}) a constant value of
$\De \chi^2$ traces out an ellipsoid in parameter space as illustrated
in figure~\ref{fig:hessecontour}a.
%
%
When the directions of the major axes of this ellipsoid coincide with
the directions of the parameter coordinate system the Hessian matrix
is diagonal and the parameters are said to be uncorrelated. If this is
not the case the Hessian can be made diagonal by a coordinate
transformation to the major axes, that is, by a rotation in parameter
space around the centre $\bi{p}_0$ of the ellipsoid. This
transformation can be written as
\begin{equation}\label{eq:uijmat}
  \De y_i = \sum_j U_{ij}\;\De p_j \qquad \sum_i U_{ij}\;U_{ik} = \de_{jk}
\end{equation}
where the second equation states that $\bi{U}$ is an orthonormal
transformation (rotation).  The matrix $\bi{U}$ is given by the
complete set of eigenvectors of the Hessian matrix as defined by the
eigenvalue equation
\begin{equation}\label{eq:eigenequation}
  \sum_j H_{ij}\;U_{kj} = \eps_{k}\;U_{ki}.
\end{equation}

The errors on the transformed parameters $y_i$ are given by $1/\sqrt
\eps_i$ so that all eigenvalues of $\bi{H}$ must be positive, which is
another way to state that the Hessian is positive definite. We mention
at this point that a non-positive definite Hessian encountered in a
(QCD) fit is a sign of either numerical problems in the calculation of
$\chi^2$ (no smooth behaviour around the minimum) or of large
correlations between the fitted parameters ($\bi{H}$ cannot be
inverted). In the latter case one or more parameters should be kept
fixed in the fit or a different parameterisation of the theory
prediction should be considered.

Rescaling $z_i = y_i \sqrt \eps_i$ maps the ellipsoid on a hyper-sphere in
$z$-space as indicated in figure~\ref{fig:hessecontour}b. Notice that
if $F$ and $G$ are given as functions of $\bi{y}$ or $\bi{z}$, instead
of $\bi{p}$, equation (\ref{eq:fgmat}) transforms to the expression
\begin{equation}\label{eq:fgmattransformed}  
  \avg{\De F \De G} = \De \chi^2
  \sum_{i} \frac{\pa F(\bi{y})}{\pa y_i}\;\frac{1}{\eps_i}\;
  \frac{\pa G(\bi{y})}{\pa y_i} = \De \chi^2
  \sum_{i} \frac{\pa F(\bi{z})}{\pa z_i}\;\frac{\pa G(\bi{z})}{\pa z_i}
\end{equation}
which is of course easier to compute than (\ref{eq:fgmat}) and,
perhaps, is also numerically more accurate.

\begin{figure}[bth]
\begin{center}
\includegraphics[width=0.8\linewidth]{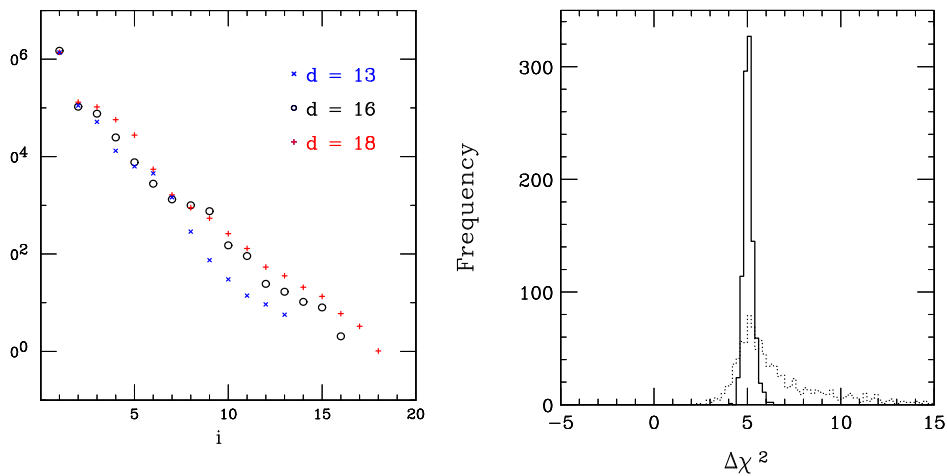}
\end{center}
\caption{\label{fig:hessecheck}
Left: The eigenvalue spectrum of the Hessian matrix from a typical
global QCD fit with (13,16,18) free parameters. Taken
from~\cite{ref:hmethod}. Right: Distribution of $\De \chi^2$
calculated by \minuit\ (dashed histogram) and by an improved calculation
from~\cite{ref:cteqpap1} (full histogram) on a 10-dimensional
ellipsoid defined by $\De \chi^2 = 5$. The spread is caused by
numerical errors in the calculation of the Hessian matrix.
}
\end{figure}

The spectrum of eigenvalues obtained from a typical QCD
fit~\cite{ref:hmethod} is shown in the left-hand plot of
figure~\ref{fig:hessecheck}.
%
%
Large (small) values of $\eps_i$ correspond to an accurate
(inaccurate) determination of the transformed parameters $y_i$ by the
fit. The large range in eigenvalues ($\approx 10^6$) implies that the
numerical calculation of the Hessian matrix must be carried out with
due attention to rounding errors. This is illustrated in the
right-hand plot of figure~\ref{fig:hessecheck} which shows the
distribution of $\De \chi^2$ calculated from the Hessian by
\minuit~\cite{ref:minuit} (dashed histogram) and by an  improved
algorithm~\cite{ref:cteqpap1} (full histogram) for parameter values
randomly distributed on the hyper-surface $\De \chi^2 = 5$.  The
improvement in the calculation of the Hessian is achieved by using
more sample points than \minuit\ in the evaluation of the second
derivatives. The algorithm is included in an update of the \minuit\ code
which can be obtained from the authors of~\cite{ref:cteqpap1}.

 
\section{Calculation of $\bi{\chi^2}$}\label{se:calculatechi}

Minimising $\chi^2$ defined by (\ref{eq:chi2def}) is impractical
because it involves the inversion of the measurement covariance matrix
(\ref{eq:vmeasdef}) which, in global fits, tends to become very
large. Because the systematic errors of different data sets are in
general uncorrelated (but not always, see~\cite{ref:nmcf2}) this
matrix takes a block diagonal form and each block could, in principle,
be inverted once and for all.  However, the dimension of these block
matrices can still easily be larger than a few hundred. Furthermore,
if the systematic errors dominate, the covariance matrix might,
numerically, be uncomfortably close to a matrix with the simple
structure $V_{ij} = \De_i \De_j$, which is singular.

Fortunately, the $\chi^2$ of (\ref{eq:chi2def}) can be cast in an
alternative form which avoids the inversion of large matrices
(we refer to \cite{ref:lmethod} for a derivation):
\begin{equation}\label{eq:cteqchi}
  \eqalign{
  \chi^2 &= \sum_i \left( \frac{m_i-t_i}{\si_i} \right)^2 
  - \bi{B}\;\bi{A}^{-1}\;\bi{B} \\
  B_k &= \sum_i \De_{ik} (m_i-t_i)/\si_i^2     \\
  A_{kl} &=  \de_{kl}
             + \sum_i \De_{ik} \De_{il} / \si_i^2. 
  }
\end{equation}
The matrix $\bi{A}$ in (\ref{eq:cteqchi}) has the dimension of the
number of systematic sources only and can be inverted at the
initialisation phase of a fitting program once the number of data
points included in the fit (i.e.\ after cuts) is known. An example of
a global QCD fit with error calculations based on the covariance
matrix approach can be found in~\cite{ref:alekhinfit}.

It is remarkable that minimising (\ref{eq:cteqchi}) is equivalent to a
fit where {\em both} the parameters $\bi{p}$ and  $\bi{s}$ are left
free. In such a fit $\chi^2$ is defined as follows. First, the
effect of the systematic errors is incorporated in the model prediction
\begin{equation}\label{eq:fpsdef}
  f_i(\bi{p},\bi{s}) =  t_i(\bi{p}) + \sum_k s_k \De_{ik}.
\end{equation}
Next, $\chi^2$ is defined by
\begin{equation}\label{eq:xpsdef}
  \chi^2 = \sum_i \left( \frac{m_i-f_i(\bi{p},\bi{s})}{\si_i} \right)^2 
  + \sum_k s_k^2.   
\end{equation}
The second term in (\ref{eq:xpsdef}) serves to constrain the fitted
values of $\bi{s}$. The presence of this term is easily understood if
one takes the view that the calibration of each experiment yields a set
of `measurements' $s_k = 0 \pm 1$~\cite{ref:agostini}.

Because $f$ is linear in $\bi{s}$ the minimisation with respect to the
systematic parameters can be done analytically. It is easy to show, by
solving the equations $\pa \chi^2 / \pa ds_k = 0$, that this leads to
the $\chi^2$ given by (\ref{eq:cteqchi}) which, in turn, is
equivalent to (\ref{eq:chi2def}), see~\cite{ref:lmethod}. The relation
between the optimal values of $\bi{s}$, the matrix $\bi{A}$ and the
vector $\bi{B}$ of (\ref{eq:cteqchi}) is
\begin{equation}\label{eq:bestsys}
  \bi{s} = \bi{A}^{-1} \bi{B}.
\end{equation}
A recent QCD analysis by the H1 collaboration~\cite{ref:h1fit} is
based on a minimisation of (\ref{eq:xpsdef}) with the systematic
parameters left free in the fit.

In the following we will use the term `Hessian method' to refer
to QCD fits based on the minimisation of the $\chi^2$ defined by
(\ref{eq:cteqchi}) or (\ref{eq:xpsdef}), the alternative being the 
`offset method' which we will describe in the next section.


\section{Offset Method}\label{se:offsetmethod}

There is another method to propagate the systematic errors which also
has the property that the inversion of a large measurement covariance
matrix is avoided. Like in the previous section $\chi^2$ is
defined by (\ref{eq:fpsdef}) and (\ref{eq:xpsdef}) but now the
systematic parameters are kept fixed to $\bi{s} = 0$ in the fit. 
This results in minimising
\begin{equation}\label{eq:offsetchi}
 \chi^2 = \sum_i \left( \frac{m_i-t_i(\bi{p})}{\si_i} \right)^2   
\end{equation}
where only statistical errors are taken into account to get the best
value $\bi{p}_0$ of the parameters.  Because systematic errors are
ignored in the $\chi^2$ such a fit forces the theory prediction to be
as close as possible to the data.

The systematic errors on $\bi{p}$ are estimated from fits where each
systematic parameter $s_k$ is offset by its assumed error ($\pm1$)
after which the resulting deviations $\De \bi{p}$ are added in
quadrature.  To first order this lengthy procedure can be replaced by
a calculation of two Hessian matrices $\bi{M}$ and $\bi{C}$
\begin{equation}\label{eq:mcijdef}
  M_{ij} = \frac{1}{2} \frac{\pa^2 \chi^2}{\pa p_i \pa p_j} \qquad 
  C_{ij} = \frac{1}{2} \frac{\pa^2 \chi^2}{\pa p_i \pa s_j}.
\end{equation}
The statistical covariance matrix of the fitted parameters is then
given by
\begin{equation}\label{eq:cstatdef}
  \bi{V}_{\rm stat} = \bi{M}^{-1}
\end{equation}
while a systematic covariance matrix can be defined by~\cite{ref:pascaud}
\begin{equation}\label{eq:csysdef}
  \bi{V}_{\rm syst} = \bi{M}^{-1} \bi{C C}^T \bi{M}^{-1}
\end{equation}
where $\bi{C}^T$ is the transpose of $\bi{C}$.  The total covariance
matrix $\bi{V}_p$ is given by the sum of the matrices $\bi{V}_{\rm
stat}$ and $\bi{V}_{\rm syst}$.

Comparing equations (\ref{eq:cteqchi}) or (\ref{eq:xpsdef})  and
(\ref{eq:offsetchi}) it is clear that the parameter values obtained by
the Hessian and offset methods will, in general, be different. This
difference is accounted for by the difference in the error estimates,
those of the offset method being larger in most cases. In statistical
language this means that the parameter estimation of the offset method
is not {\em efficient}. The offset method has a further disadvantage
that the goodness of fit cannot be judged from the $\chi^2$ which is
calculated from statistical errors only.  An {\it ad hoc} solution to
this problem is to re-calculate, after the fit has converged, a
$\chi^2$ with the statistical and systematic errors added in
quadrature~\cite{ref:mbfit}.

\begin{figure}[tbh]
\begin{center}
\includegraphics[width=0.8\linewidth,height=7cm]{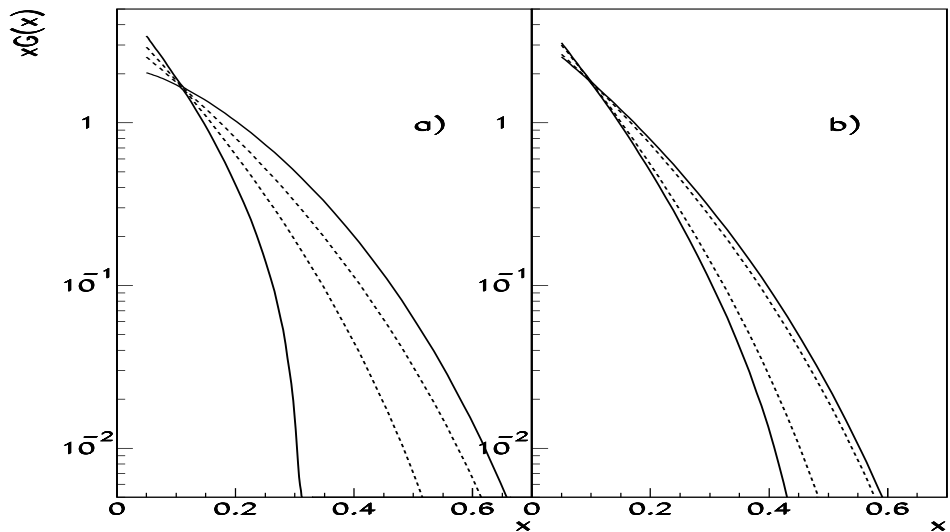}
\end{center}
\caption{\label{fig:alekhinglue}
The gluon density versus the Bjorken scaling variable $x$ from
a LO QCD fit to the BCDMS structure function data. The dotted curves show the
statistical error band. The full curves indicate the systematic error
band calculated using the offset method (a) and the Hessian method
(b). Figure taken from~\cite{ref:alekhinstudy}.
}
\end{figure}

For a detailed comparison of the Hessian and offset methods we refer
to~\cite{ref:alekhinstudy} where it is shown that the error estimates
from the two methods can differ by a large amount when the systematic
errors dominate.  This is illustrated in figure \ref{fig:alekhinglue}
which shows the error bands on the gluon density obtained from a LO
QCD fit to the BCDMS data.
%
%

This might also explain the difference in the error estimates
on $\alpha_s$ from recent QCD fits by H1~\cite{ref:h1fit} (Hessian
method with free systematic parameters) and ZEUS~\cite{ref:zeusfit}
(offset method):
\[
  \alpha_s(M_Z^2) = \left\{
  \begin{array}{ll}
   0.1150 \pm 0.0017\ {\rm (exp.)} & {\rm H1} \\
   0.1172 \pm 0.0055\ {\rm (exp.)} & {\rm ZEUS\ preliminary.}
  \end{array}
  \right.
\]
Notice however that these fits differ in many other respects like the
data sets included, kinematic cuts, treatment of charm mass effects
and so on.


\section{Exploring the $\bi{\chi^2}$ Profile}\label{se:lagrangian}

As mentioned above, the Hessian method is based on the assumption that
the theory prediction is approximately linear in the vicinity of
$\bi{p}_0$ which means that $\chi^2$ is a quadratic function of the
parameters near the minimum. To check this quadratic dependence one
can fix a parameter $p_i$, say, and optimise the remaining parameters
for different input values of $p_i$.  In this way $\chi^2$ is explored
along the axes of the parameter coordinate system. The procedure is
automatically carried out using the `Minos' option in \minuit.

The Lagrange multiplier method, developed in~\cite{ref:lmethod},
allows to investigate the $\chi^2$ profile along any relevant
direction in parameter space. Here the quantity
\begin{equation}\label{eq:lagrangechi}
  \chi^2(\bi{p},\la) = \chi^2_g(\bi{p}) + \la X(\bi{p})
\end{equation}
is minimised for several fixed values of the Lagrange multiplier
$\la$. In (\ref{eq:lagrangechi}) $\chi^2_g$ is the global $\chi^2$
calculated from the data included in the fit and $X$ is a physics
quantity of interest ({\em not} included in the fit), for instance,
the $W$ production cross-section $\si_W$ in $p\bar{p}$ collisions at
the Tevatron. The results of such a lengthy analysis obtains the
$\chi^2$ profile as a function of $X$ and thus the range of $X$
corresponding to a given value of $\De \chi^2$. This method does not
make use of a quadratic approximation of the $\chi^2$ profile.

\begin{figure}[tbh]
\begin{minipage}[t]{0.47\textwidth}
\centering
\includegraphics[width=0.8\linewidth]{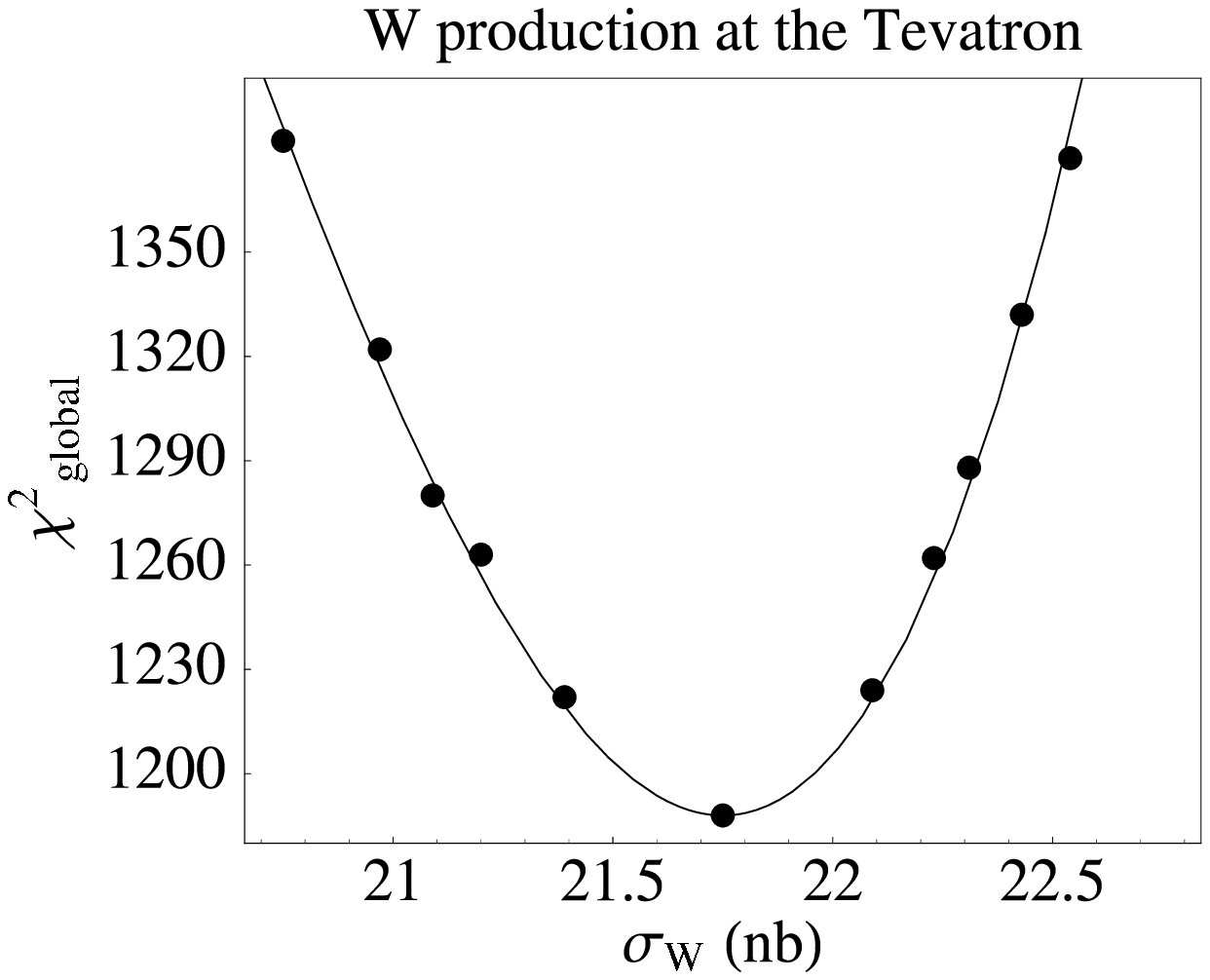}
\end{minipage}
\begin{minipage}[t]{0.53\textwidth}
\centering
\raisebox{7mm}{
\includegraphics[width=0.8\linewidth]{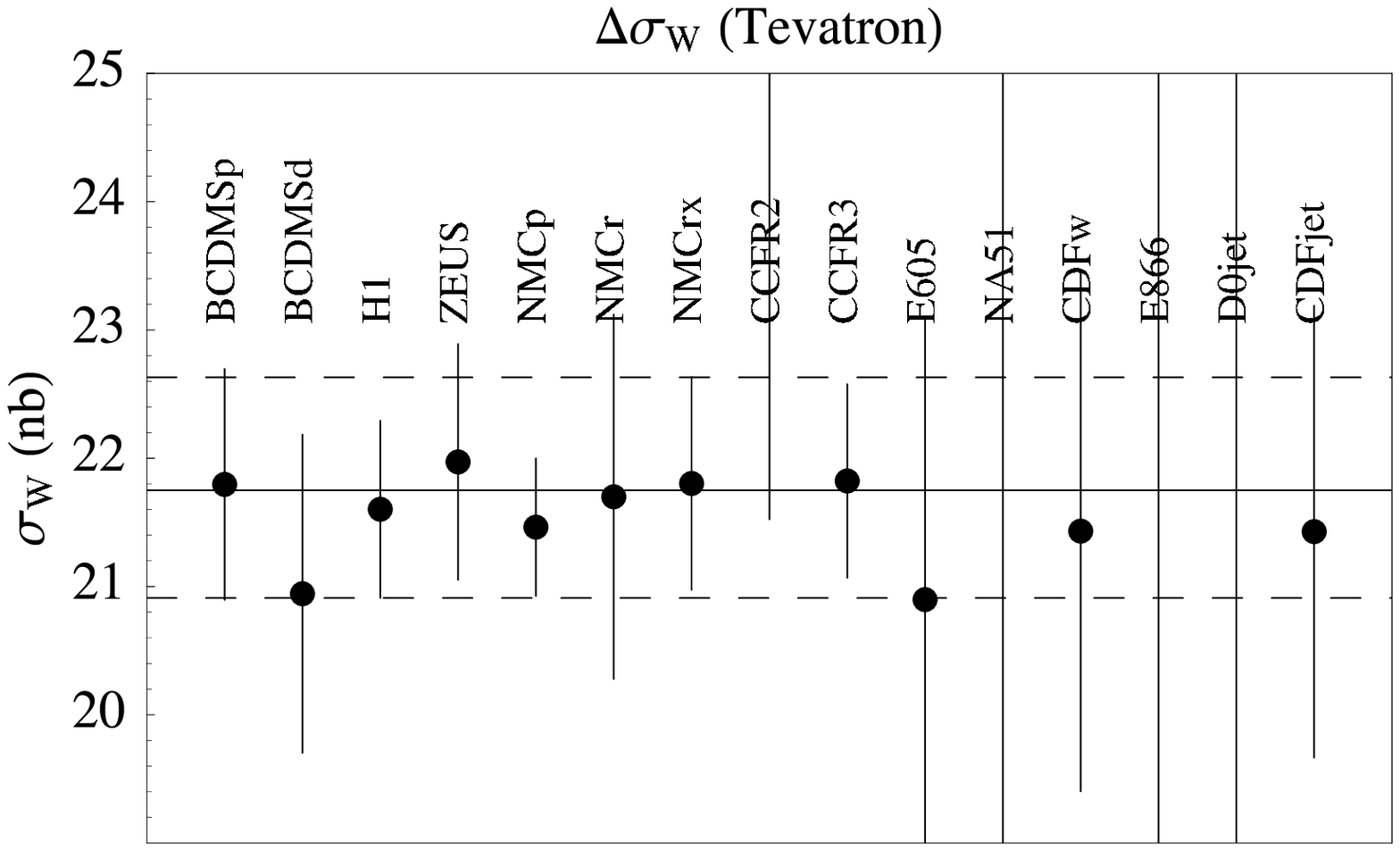}
}
\end{minipage}
\caption{\label{fig:lagrange}
Left: The $\chi^2_g$ (see text) versus the cross-section $\si_W$ of
inclusive $W$ production at Tevatron energies from the QCD analysis
of~\cite{ref:lmethod}. Right: Optimal values of $\si_W$ (dots) and
90\% confidence levels (bars) for each data set included in the fit.
The full line shows the best prediction for $\si_W$ and the dashed
lines represent the confidence bounds described in the text. Figures
taken from~\cite{ref:lmethod}.
}
\end{figure}

In the left-hand plot of figure~\ref{fig:lagrange} we show $\chi^2$
as function of $\si_W$ from the analysis of~\cite{ref:lmethod}.
%
%
In this analysis the 90\% confidence levels of $\si_W$ were obtained
from the $\chi^2$ profiles of each data set individually, see the
right-hand plot of figure~\ref{fig:lagrange}. The uncertainty on
$\si_W$ was then defined as the intersection of these individual
confidence levels (dashed lines in figure~\ref{fig:lagrange}) giving $
20.9 < \si_W < 22.6$~nb.  A similar uncertainty of 4\% on the $\si_W$
prediction is obtained  from the Hessian method provided $\De \chi^2$
in (\ref{eq:hessian}) or (\ref{eq:fgmat}) is set to 180 (for a fit
of 1295 data points distributed over 15 data sets).

The origin of this large $\De \chi^2$ is unclear to us but the
question which value of $\De \chi^2$ should be chosen in a global QCD
analysis and which deviations from the expected $\chi^2$ value ($N \pm
\sqrt{2N}$ for a fit with $N$ degrees of freedom) can be  tolerated is
clearly an important issue. For a discussion on this subject we refer
to~\cite{ref:collins}.

A bad $\chi^2$ in a global analysis may have several causes. First, it
can be an indication of physics beyond the Standard Model. Second, the
theoretical modelling may be inadequate because higher order terms in
the perturbative expansion are missing or non-perturbative
contributions like higher twists or nuclear effects are not, or only
partially, taken into account. In addition, there is always the
question if the parton densities are parameterised with sufficient
flexibility.  Third, the information on the experimental errors may be
inaccurate, incomplete (not all correlations given) or even not be
fully known. Finally, the data may very well not be Gaussian
distributed.

Concerning the latter point we refer to an
analysis~\cite{ref:bukhvostov} of a large sample of data from the
Table of Particle Properties. It turns out that the probability
distribution of this body of data is far from Gaussian. This may be
due to uncertainties in the error estimates provided by the
experiments which, as is shown in~\cite{ref:bukhvostov}, can strongly
affect the shape of the probability  distribution of the data.


\section{Parton Distribution Sets}\label{se:pdfset}

Error calculations in global QCD fits are of little practical use if
the results are not made available in the form of parton distribution
sets which contain the full information on uncertainties and
correlations. To our knowledge two such sets exist at present.

The set provided by Alekhin~\cite{ref:alekhinfit} gives as a function
of $x$ and $Q^2$ the values of the parton densities and their
covariance matrix as calculated with~(\ref{eq:fgmat}). This allows to
compute the error on any function of the parton distributions but
only at a given kinematic point. It is, for instance, not possible to
evaluate the errors on (convolution) integrals since the information
on the correlation between different kinematic points is lost. Notice
that the errors from~\cite{ref:alekhinfit} are defined by the Hessian
Method described in section~\ref{se:calculatechi}.

The \epdflib\ set~\cite{ref:epdflib} based on the QCD analysis
of~\cite{ref:mbfit} gives the covariance matrix of the fitted
parameters (from the offset method described in
section~\ref{se:offsetmethod}) and, as functions of $x$ and $Q^2$, the
parton densities as well as their derivatives to all the fitted
parameters.\footnote{The \epdflib\ set is available from
{\tt http://www.nikhef.nl/user/h24/qcdnum.}}
 From this information the error on any function of the
parton densities can be calculated with~(\ref{eq:fgmat}).  As an
example let us consider a hadron-hadron cross section which can be
written as a convolution of the parton densities and a hard scattering
cross section, generically,
\begin{equation}\label{eq:hard}
\si = \sum_{ij} f_i \otimes f_j \otimes \hat{\si}_{ij}.
\end{equation}
To calculate the error on $\si$ with (\ref{eq:fgmat}) it is sufficient
to compute the derivatives
\begin{equation}\label{eq:sighadron}
\frac{\pa \si}{\pa \pla} = \sum_{ij} \left[
\frac{\pa f_i}{\pa \pla} \otimes f_j +
f_i \otimes \frac{\pa f_j}{\pa \pla} \right] \otimes \hat{\si}_{ij}
\end{equation}
which is straight forward since both $f_i$ and the derivatives are
available from \epdflib. A practical example of the use of \epdflib\ in
an analysis of dijet production at HERA can be found
in~\cite{ref:zeusdijet}.

\begin{figure}[tbh]
\begin{center}
\includegraphics[width=0.8\linewidth]{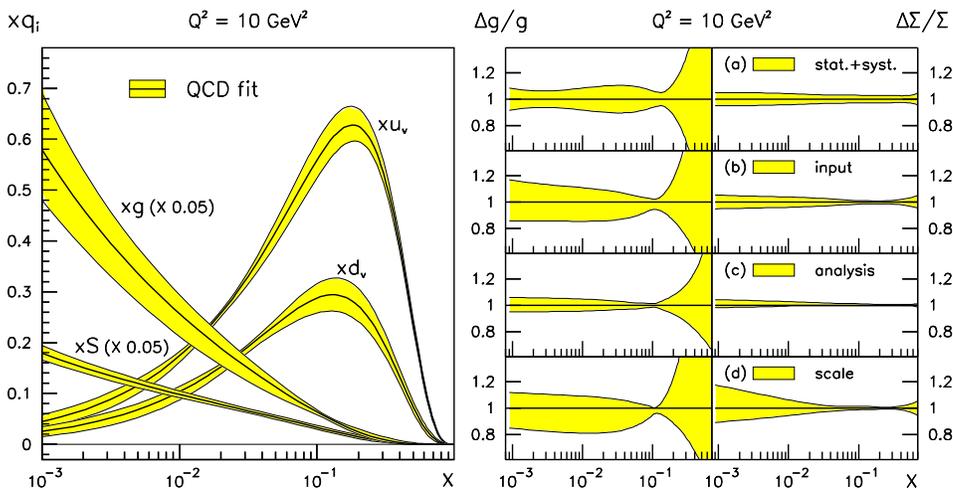}
\end{center}
\caption{\label{fig:errorbands}
Left: The parton momentum densities from the QCD analysis
of~\cite{ref:mbfit} versus $x$ at $Q^2 = 10$~GeV$^2$. Right: The
errors on the gluon and quark densities from the various sources
described in the text.  Figures taken from~\cite{ref:mbfit}.
}
\end{figure}

In figure~\ref{fig:errorbands} we show the parton densities from
\epdflib\ (left hand plot) and the relative error contributions to the
gluon and singlet quark densities (right hand plot).
%
%
The \epdflib\ set provides, in addition to the experimental
statistical and systematic errors, information on the following
sources of uncertainty:
\begin{itemize}
\item Uncertainties due to those on the input parameters of the
  analysis like $\alpha_s$, heavy flavour thresholds, nuclear effects
  and so on. Parton densities are provided with each of these input
  parameters offset by their errors. These densities can be used to
  either define an error band or to obtain, by interpolation, densities
  with varying input conditions;
\item Analysis error defined as the envelope of the central fit and 10
  alternative fits (vary cuts, input scale etc.) which all gave
  acceptable values of $\chi^2$. This error band quantifies the
  stability of the QCD fit;
\item Parton densities obtained from fits where the renormalisation and
  factorisation scales were independently varied in the range
  $Q^2/2 < \mu^2_{R,F} < 2Q^2$. Again, these densities can be used to
  define an error band or be interpolated to obtain the distributions
  for a particular choice of scale.
\end{itemize}

A simple but important check is provided by the calculation of the
uncertainty on the total momentum fraction carried by quarks and
gluons. This error should vanish because the momentum sum was
constrained to unity in the analysis of~\cite{ref:mbfit}. Indeed we
find with \epdflib\ for the values and errors of these momentum fractions
at $Q^2 = 4$~GeV$^2$
\begin{eqnarray}\label{eq:srule}
\int_{0.001}^1 xg \, dx \hspace{13mm} & = & 0.393 \pm 0.018 
\ \ \mathrm{(stat. + syst.)} \nonumber \\ 
\int_{0.001}^1 x\Si \, dx \hspace{12mm} & = & 0.594 \pm 0.018 \nonumber \\
\int_{0.001}^1 (xg + x\Si)\, dx & = & 0.987 \pm 0.002 \nonumber 
\end{eqnarray}
where the error on the last integral is much smaller than that on the
first two, as it should be. This example clearly illustrates the
importance of taking into account the correlations between the errors
on the parton densities.


\section{Summary}\label{se:summary}

In this report we have presented an overview of the least squares
minimisation and error propagation techniques used in  many recent
global QCD fits.  The aim of these global analyses is to determine
from a large and diverse body of scattering data the parton density
distributions as well as their errors and correlations.

Assuming that the measurement errors are Gaussian distributed the
likelihood function can be written as a multivariate Gaussian
distribution. This leads to a $\chi^2$ definition which can have
different, but mathematically equivalent representations.  In
particular it turns out that a fit using the full covariance matrix of
the data is equivalent to a fit where the systematic correlations are
included in the model prediction together with the introduction of a
set of free systematic parameters.

Error propagation based on shifting the data by the systematic errors
and adding the deviations in quadrature is {\em not}
equivalent to the method described above and leads to theory
predictions which are as close as possible to the fitted data at the 
expense of larger error estimates, in particular when the systematic
uncertainties dominate.

Several technical issues are addressed such as the numerical
accuracy of the calculation of the Hessian matrix, the Lagrange
multiplier method to explore the multi-dimensional $\chi^2$ profile in
some physically relevant direction and the representation of the
global QCD fit results in publically available parton distribution
sets which contain the full information on errors and correlations.

I am grateful to D.~Stump for providing me with mathematical proofs of
the equivalence of several $\chi^2$ representations and to
S.~Alekhin, J.~Pumplin, W.K.~Tung and A.~Vogt for  discussions and
comments on the manuscript. I thank the organisers for inviting me to
an excellent and stimulating workshop.

\section*{references}

\end{document}